# Ice-templated poly(vinylidene fluoride) ferroelectrets

Yan Zhang[1], Chris R. Bowen[1], Sylvain Deville[2]

[1] Department of Mechanical Engineering, University of Bath, Bath, UK
[2] Laboratoire de Synthèse et Fonctionnalisation des Céramiques, UMR 3080 CNRS/Saint-Gobain CREE, Cavaillon, France

**Abstract**

Ferroelectrets are piezoelectrically-active polymer foams that can convert externally applied loads into electric charge. Existing processing routes used to create pores of the desired geometry and degree of alignment appropriate for ferroelectrets are based on complex mechanical stretching and chemical dissolution steps. As a simple, cost effective and environmentally friendly approach, freeze casting is able to produce aligned pores with almost all types of the materials, including polymers. In this work, we present the first demonstration of freeze casting to create polymeric ferroelectrets. The pore morphology, phase analysis, relative permittivity and direct piezoelectric charge coefficient ($d_{33}$) of porous poly(vinylidene fluoride (PVDF) ferroelectrets with porosity volume fractions ranging from 24% to 78% were analysed. The long-range alignment of pore channels produced during directional freezing was shown to be beneficial in forming a highly polarised structure after breakdown of air in the pore channels via corona poling. This new method opens a way to create tailored pores and voids in ferroelectret materials for transducer applications related to sensors and vibration energy harvesting.

**Introduction**

Piezoelectric materials manifest themselves by the generation of a charge when subjected to mechanical loads. The direct piezoelectric effect can be used to sense dynamic pressure, acceleration, or change in force and there is also potential for scavenging energy from motion in the surrounding environment [1]. *Ferroelectric* materials are a sub-class of piezoelectric materials and exhibit piezoelectric properties as a result of a remnant polarisation due to the presence of aligned domains, which have been of interest in sensor and energy harvesting applications [2]. In order to evaluate the performance of such materials, relevant figures of merit such as $d_{ij}^2/\varepsilon_{33}^\sigma$ for piezoelectric energy harvesting, while the piezoelectric voltage coefficient, such as $g_{33}=d_{33}/\varepsilon_{33}^\sigma$ for sensor applications have been widely used, where $d_{ij}$ is the piezoelectric charge coefficient, and $\varepsilon_{33}^\sigma$ is the permittivity of the material at constant stress. These figures of merit indicate that a low permittivity, and high piezoelectric activity are beneficial for sensor and energy during harvesting applications. Ferroelectric ceramics often exhibit high figures of merit for high sensor sensitivity or high energy capability, but they are relatively high density and exhibit poor mechanical flexibility. Ferroelectric polymers exhibit a relatively low piezoelectric activity, but they have a low permittivity, are lightweight and exhibit flexibility which is important for flexible electronics, wearables and stretchable electronics [3].

As an alternative to ferroelectric materials, *ferroelectrets* are a class of piezoelectrically-active foam manufactured from a non-polar porous polymer whereby gas, such as air, within the pore space can be subjected to electrical breakdown when subjected to a high electric field during a poling process. This process results in opposing electric charges being deposited on the upper and lower surfaces of the pores [4]. As a result of a poling process, a dipole-like structure is formed where the dipole moment can be changed by applying a mechanical stress, thereby leading to a piezoelectric response. The piezoelectric $d_{33}$ charge coefficient, a measure of the charge generated per unit force, can be an order of magnitude greater than those found in conventional ferroelectric polymers, such as polyvinylidene fluoride (PVDF) and can be comparable or even greater than ferroelectric ceramics [5, 6] while maintaining a high compliance due to their polymeric nature. For example, for sensing applications a ferroelectret polymer exhibited a $g_{33}$, a measure of the electric field per unit stress, that was 10-150 times higher than ferroelectric PVDF and over a thousand times higher than that of lead zirconate titanate (PZT), due to their high $d_{33}$ values (e.g. 25-700 pC/N for porous polypropylene) and low permittivity (e.g. $\varepsilon_{33}^\sigma$ =1.12-1.23 for porous polypropylene).[7, 8] Ferroelectret polymers with an average pore diameter of 0.9 µm or a pore height of 4.5 µm and pore width of 1.55 µm were also demonstrated to produce 44.9 nW [9] and 4.35 mW [10] for energy harvesting applications. Therefore, ferroelectrets have gained much interest in low-level mechanical energy harvesting [11] and sensing [4] based on their relatively high piezoelectric coefficient and low permittivity.

To create ferroelectret materials with a high electro-mechanical response, a cellular geometry or lens-shaped pore structure is desirable with pores that are elongated in the lateral extension; this can be typically > 10 µm long and only a few µm in height [12, 13]. Such a morphology is beneficial due to the relatively low elastic stiffness of such an anisotropic pore structure in the polarisation direction [14, 15]. Micron-size pores are desirable to achieve micro-discharges within pores, which are interpreted in terms of Paschen breakdown [16]; and large electric fields are required in very small pores. As a result, the majority of existing processing routes have focused on creating cellular or lens-shaped microstructures for ferroelectrets. Current processing routes include biaxial stretching of polymers with embedded foreign particles followed by chemically dissolution of the embedded foreign particles [9, 10], or the injection of the polymer by high-pressure gas to further modify the pore size and shape [17, 18].

Freeze casting, also termed the ice-templating method [19], is an effective and facile technique to prepare materials with tailored pore morphology and anisotropic porosity. This approach involves freezing a colloidal suspension under a unidirectional gradient temperature, followed by sublimation of the solvent. The aligned pore channels formed parallel to the freezing direction are direct replica of the solvent after solidification. Freeze casting has potential to be a cost effective and environmentally friendly approach, and it is feasible to tailor pore characteristics and form aligned pores in almost all types of materials: ceramics, polymers, metals and their composites [19]. To date, more than 30 different types of polymers with straight, aligned and elongated pore channels have been formed by freeze casting for catalysis, separation, biomedical and thermal insulation applications. [19] However, there has been no report to date on the freeze casting of polymers to create polymeric ferroelectrets.

In this work we used freeze casting of PVDF to produce a porous ferroelectret material. PVDF is a classical ferroelectric polymer which is partially crystalline and has the dipole moment perpendicular to the polymer chain, especially when the polar all-trans β-phase is dominant [20]. Copolymers of the materials have been employed to achieve a higher crystallinity for improved ferroelectric properties; these include poly(vinylidene fluoride-trifluoroethylene) (PVDF-TrFE) and poly(vinylidene fluoride-hexafluoropropylene) (PVDF-HFP) [21, 22]. PVDF has been selected since freeze casting has been previously employed to form a PVDF microporous membrane with pore sizes < 10 µm whose Young's modulus and water flux ability were explored [23], but no piezoelectric properties were reported. Here we have examined the fabrication of the optimised pore structure in PVDF formed by freeze casting, with proof of its ferroelectret nature and corresponding piezoelectric response. A range of polymer solutions at different polymer loading levels were assessed for freeze casting and the microstructure, phase identity, direct $d_{33}$ piezoelectric charge coefficients and permittivity of the polymers were examined in detail.

**Experimental**

Poly(vinylidene fluoride) (PVDF, 6010) powders were used from Solef® (Belgium) and used in the as-received condition. A schematic of the freeze casting process applied to PVDF is shown in Fig. 1. PVDF powders with different weight fractions of 0.2, 0.4, 0.6, 0.8, 1 and 1.2 g were dissolved in 10 g solvent of dimethyl sulfoxide (DMSO, anhydrous≥99.9%, freezing point of 19°C, Sigma-Aldrich) and stirred at 60°C for overnight to achieve homogenous solutions; see Fig. 1(A). Other widely used polar solvents (e.g. N-methyl-2-pyrrolidone/NMP, dimethylformamide/DMF, dimethylacetamide/DMAc) with lower melting points of -20 to -61°C were not suitable for the demoulding step at room temperature. The prepared PVDF/DMSO solutions were then poured into polydimethylsiloxane (PDMS) mould with a diameter of 20 mm and height of 10 mm. The bottom face of the mould was held on the surface of liquid $N_2$ and the top face of the mould held at an ambient temperature of ~20°C to unidirectionally freeze the solution. In a conventional freeze casting process [24], freeze drying is undertaken at low temperature and pressure in order to sublimate the solvent from a solid to a gas. In this work, freeze drying of the frozen PVDF/DMSO was not chosen since DMSO has the potential to contaminate not only the vacuum pump [25], but has an unpleasant sulfur smell [26] with potential health risks during sublimation of the solvent [27]. After freezing, the samples were therefore demoulded and immediately immersed into water to dissolve and remove the solidified DMSO, which was repeated 7-8 times to ensure fully remove the DMSO. The samples were then dried in an oven at 40°C for 24hrs to eliminate the remaining water (Fig. 1B). The dried samples were then frozen in liquid $N_2$, followed by cutting to a thickness of ~1.5 mm parallel to the freeze-cast direction using a sharp blade (Fig. 1C). Materials were cut parallel to the freezing direction to pole the pores that were aligned and elongated in the direction normal to the polarisation direction. Corona poling was conducted by applying a DC voltage of 26.8 kV for 30 min at ambient temperature (Fig. 1D). All samples were silver electroded (RS Components, Product No 186-3600, UK) on both sides normal to the freeze-cast direction for piezoelectric and dielectric characterisation.

The microstructure of the samples was examined by scanning electron microscopy (SEM, JSM6480LV, Tokyo, Japan). The bulk density of the sintered specimens was measured using Archimedes' principle. The piezoelectric strain coefficient ($d_{33}$) was measured using a Berlincourt Piezometer (PM25, Take Control, UK) 0.5-6 days after corona poling. To confirm that the measured $d_{33}$ values originate from the piezoelectric effect, the sample orientation was reversed to ensure the piezoelectric coefficient changes from a positive to negative polarity. The relative permittivity (ε) and dielectric loss (tan δ) were measured in the frequency rage of 1 to $10^6$ Hz with an oscillating voltage of 1 $V_{rms}$ using an Agilent Technology 4192A impedance analyzer. Fourier transform infrared spectra (FT-IR) of the materials were recorded using a Perkin Elmer Spectrum 100 with a diamond universal ATR attachment. The phase structure of the ceramics was examined by X-ray diffraction (BRUKER D8-Advance, USA) with Cu radiation with 2θ ranging from 10° to 60°.

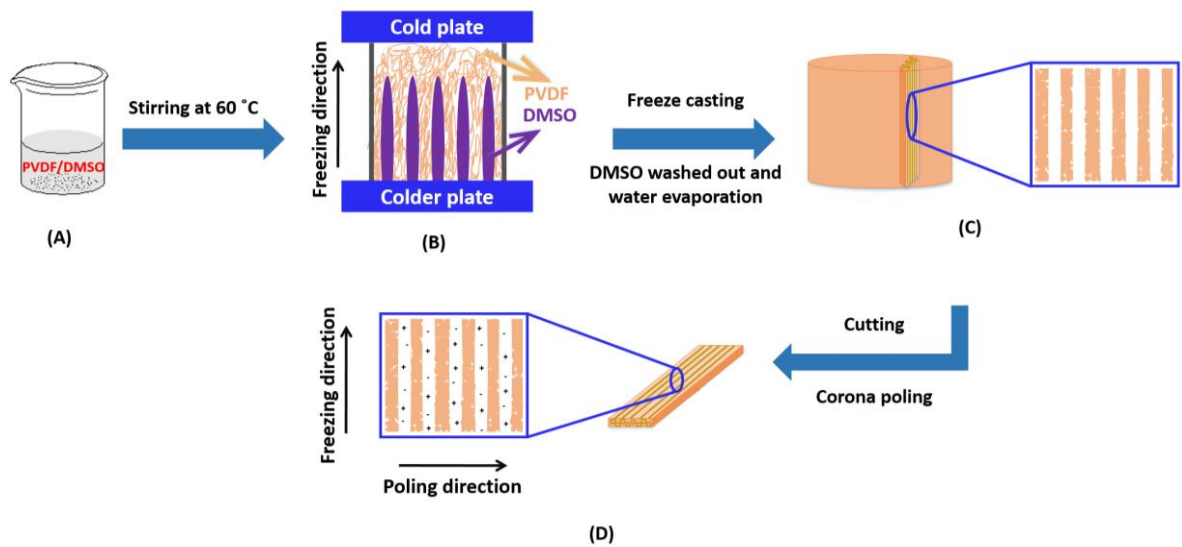

Figure 1: Schematic of porous polymer obtained from freeze casting. (A) formation of PVDF/DMSO solution, (B) the freeze casting process, (C) sample sectioning parallel to freezing direction after solvent removal, (D) corona poling of sectioned PVDF.

**Results**

Microstructural observations of the freeze cast porous PVDF materials, phase analysis, dielectric spectroscopy and piezoelectric properties of the poled freeze-cast materials are now described.

*Microstructure*

Figure 2 A–F show the porous structure of PVDF prepared from the solutions with weight ratios of PVDF/DMSO from 0.2/10 to 1.2/10. At a ratio of 0.2/10 with the minimum PVDF content, adjacent pore channels with a pore size less than 5 µm were formed that were overlapped, see Fig. 2A. A further increase in the weight ratios from 0.4/10 (Fig. 2B) to 0.8/10 (Fig. 2D) led to the formation of more aligned pore channels along the freeze-cast direction, and the pore size normal to the freeze-cast direction decreased from 25 ± 3 µm in the 0.4/10 sample to 20 ± 2 µm in the 0.8/10 sample (Fig. 2D). At low magnification aligned pores can be observed on the upper sample surface, where in Fig. 3A mm-scale length pores can be seen in the freeze-cast PVDF with a ratio of 0.4/10. A similar low magnification side view, see inset of Fig. 3B, reveals the lens-shaped pore morphology. Compared to commercially available ferroelectrets [17], the freeze-cast polymers had a larger pore length with similar pore shape and pore size which might facilitate the generation and deposition of plasma charges on the inner pore surface under the action of a high electric field during the poling process. A further increase in the amount of PVDF to a ratio of 1.0/10 (Fig. 2E) and 1.0/10 (Fig. 2F) resulted in an overlapped pore morphology (Fig. 2E) with the formation of more rounded pores, as in Fig. 2F, rather than an aligned structure; this is similar to the structure of ice-templated ceramics when the solid loading is too high [28].

In general, larger pores are formed from the PVDF/DMSO solution at low PVDF concentrations during directional freezing, since the porous structure is a replica of DMSO crystals. However, for a low PVDF content of 0.2/10, the capillary force [29] formed by water during evaporation during drying leads to shrinkage of the pore channels, thereby making the adjacent pore channels overlap, as shown in Fig. 2A. With an increase of PVDF concentration, there is a reduced effect of the capillary force on pore deformation (Fig. 2B-D) for the 0.4/10 to 0.8/10 samples, due to the lower amount of water left in the samples. However, if PVDF concentration is too high there is a loss of directionality and smaller equi-axed pores were formed; this is possibly due to the high viscosity of the solution during freeze casting [30], (Fig 2E–F).

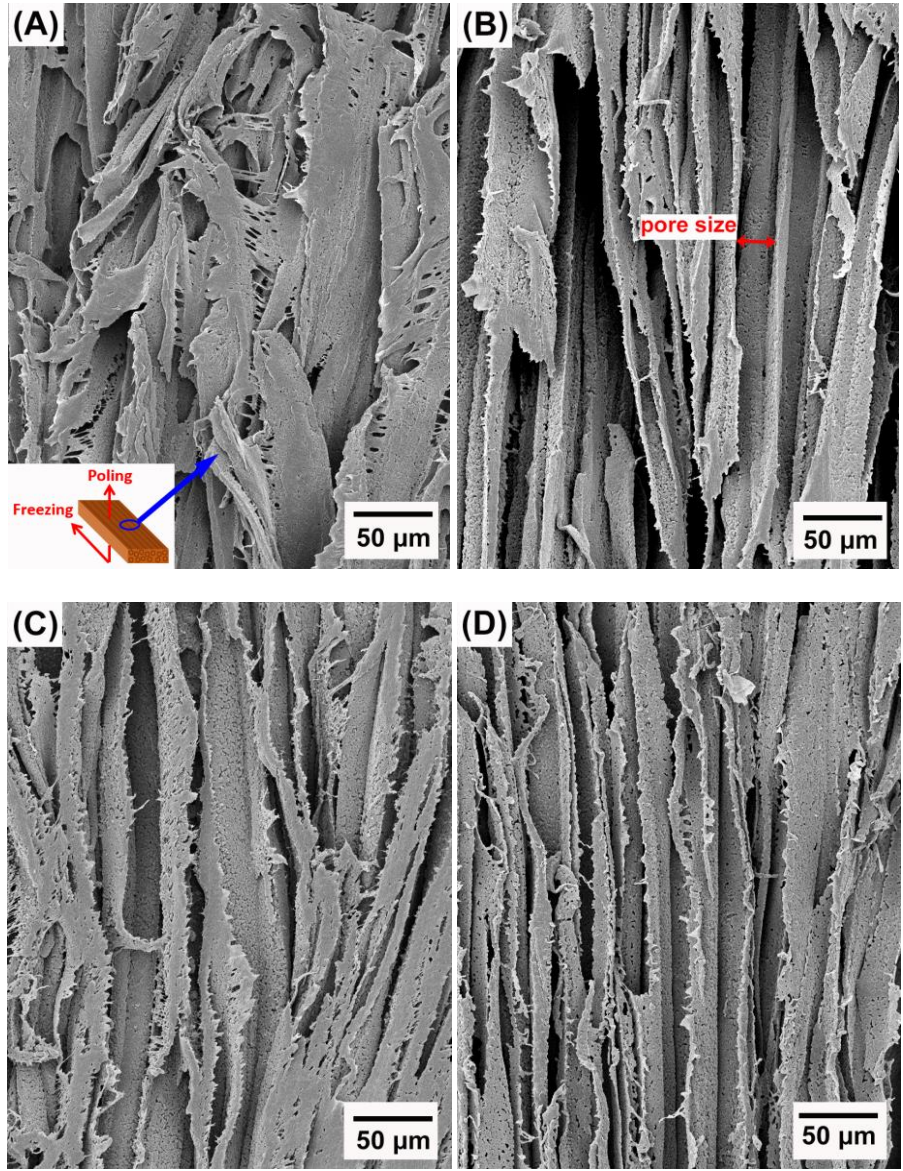

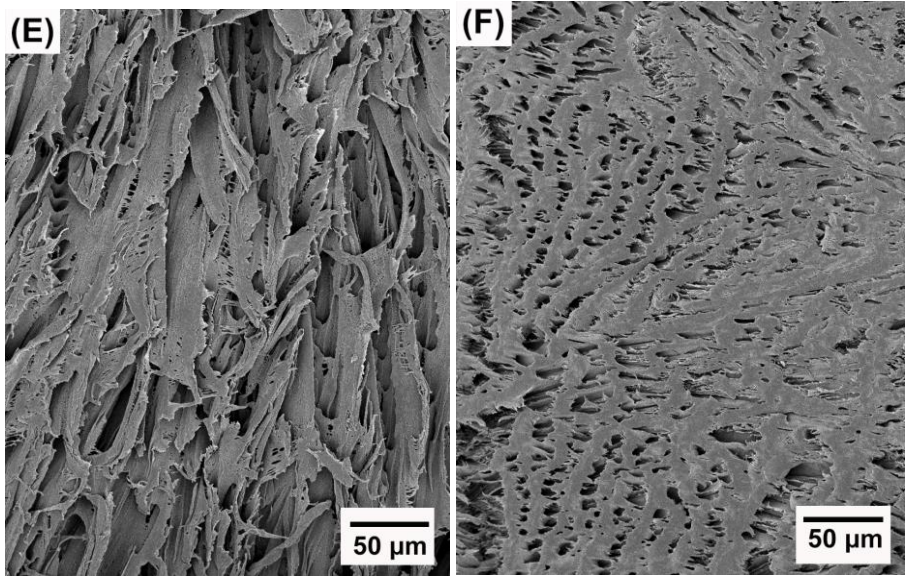

Figure 2: Scanning electron micrographs of porous freeze cast PVDF with increasing weight ratios of PVDF/DMSO. (A) 0.2/10, (B) 0.4/10, (C) 0.6/10, (D) 0.8/10, (E) 1.0/10, (F) 1.2/10.

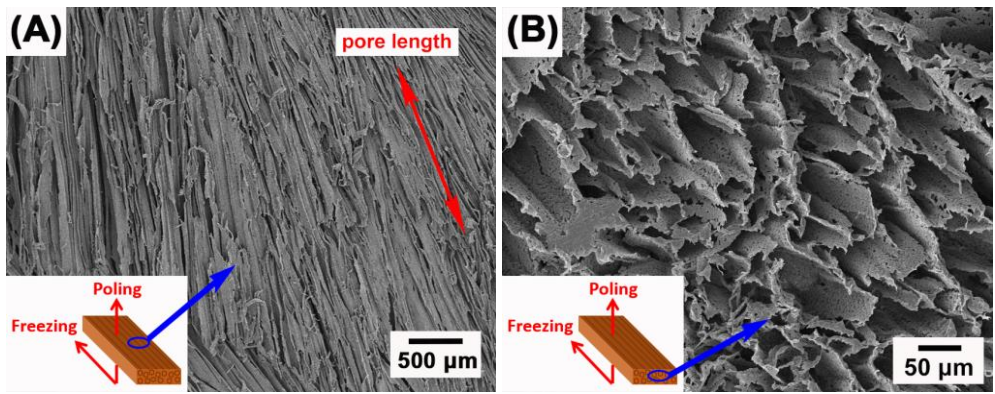

Figure 3: Scanning electron micrographs of (A) the mm-scale pore length and (B) the prism/lens pore morphology of the freeze-cast PVDF with the weight ratio of 0.4/10.

*Phase analysis*

Figure 4 shows the FT-IR and XRD spectra of the freeze cast porous PVDF with different weight ratios of PVDF/DMSO. The FT-IR transmittance spectra are shown in the 550-1500 cm$^{-1}$ (Fig. 4A), and 800-900 cm$^{-1}$ (Fig. 3B) ranges. No characteristic peak for either the α- or β- phase was found in all freeze cast samples. Some peaks could be attributed to both β- and γ-phases, namely 1176 cm$^{-1}$ (Fig. 4A) and 840 cm$^{-1}$ (Fig. 4B) [31, 32].

However, the representative peak at 1234 cm$^{-1}$ can be exclusively used to identify the γ-phase characteristic band [33, 34], which is uniquely assigned to the γ-crystalline form [35, 36]. A small peak at 811 cm$^{-1}$ was also observed, and is also a peak exclusively attributed to the γ-phase [31]. Fig. 4C and Fig. 4D show the X-ray diffraction of the polymers with different weight ratios of PVDF/DMSO. A typical semi-crystalline peak appeared in the XRD curves indicating the semi-crystalline nature of the materials for all weight ratios of PVDF/DMSO. All PVDF samples exhibited peaks centred at a diffraction angle of 20.04° (110) [33] which further confirmed the γ-phase crystal polymorph was dominant in the final materials. Therefore both infrared spectroscopy and XRD spectra in Fig. 4 indicate that the γ-phase was the predominant crystalline presence in the porous PVDF for all PVDF/DMSO weight ratios, which is in agreement with previous conclusions that γ-phase usually was normally obtained from the DMSO solvent [37, 38], regardless of the preparation temperature [39, 40].

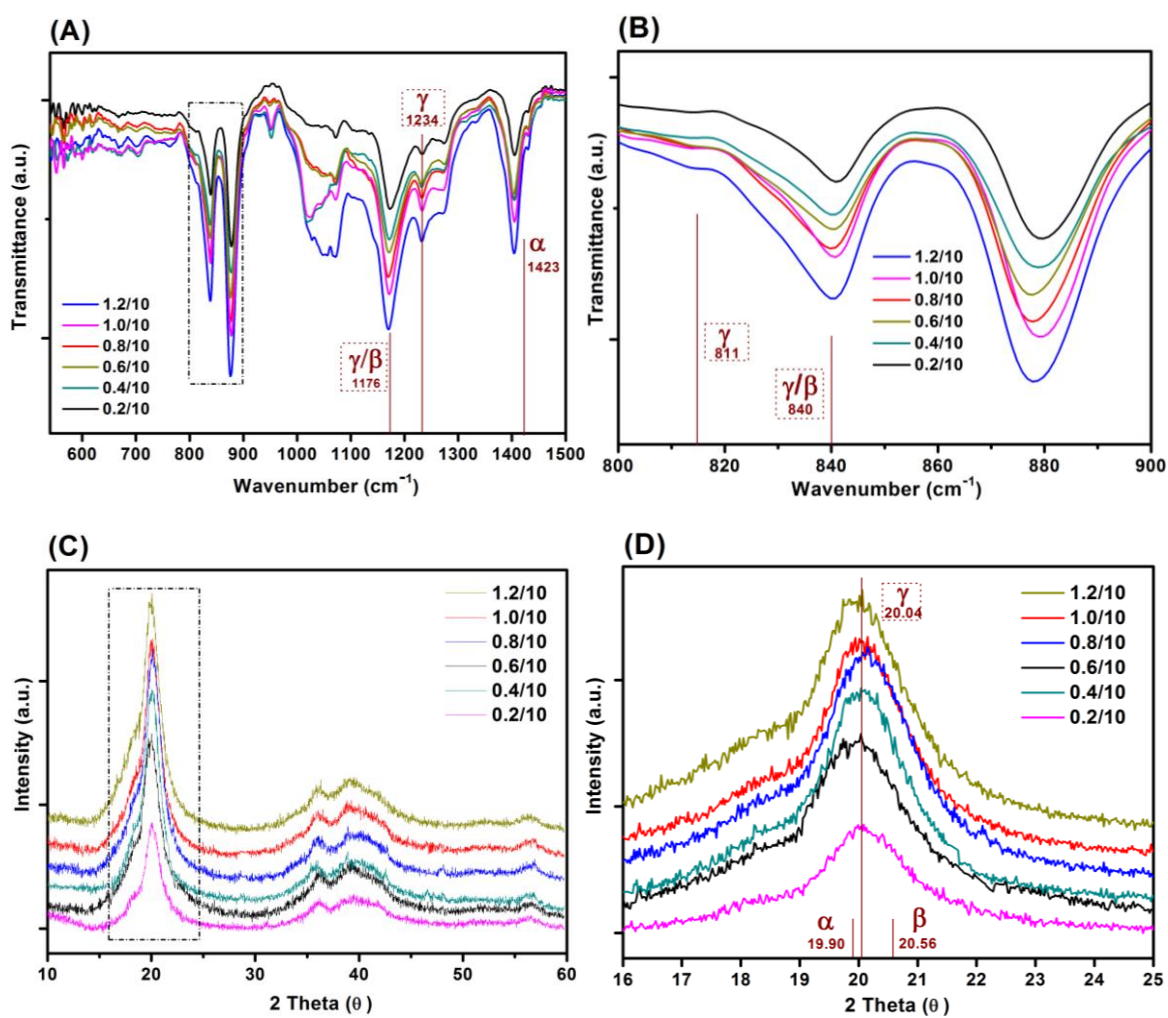

Figure 4: (A) FTIR and (B) XRD spectra of porous PVDF with different weight ratios of PVDF/DMSO, (C) and (D) are FTIR and XRD spectra in the range of 800-900 cm$^{-1}$ and 16-25°, respectively.

*Piezoelectric and frequency dependent dielectric properties*

Figure 5 A–D shows the porosity variation and piezoelectric performance of the poled porous PVDF with weight ratios of PVDF/DMSO ranging from 0.2/10 to 1.2/10. The apparent porosity increased from 35 vol.% for 0.2/10 sample to reach the maximum value of 78 vol.% for 0.4/10, followed by a decrease to 24 vol.% for 1.2/10 (Fig. 5A). The higher porosity level of the aligned porous structure may provide more space to retain the plasma charges for piezoelectric activity. As the PVDF/DMSO weight ratio increased from 0.2/10 to 0.8/10, the $d_{33}$ values increased from 41 ± 1 pC/N for 0.2/10 to 264 ± 2 pC/N for 0.4/10; this was followed by a decrease in $d_{33}$ for ratios higher than 0.6/10; see Fig. 5B. The $d_{33}$ values obtained for these materials were comparable with ferroelectrets prepared from stretching and gas injection methods [5, 6, 41], and a commercial ferroelectret with a reported values of 25 pC/N to 200 pC/N [42]. In addition, Fig. 5B clearly shows the almost equal in magnitude and opposite charge accumulations at the two-opposing poled-sides in terms of negatively-poled and positively-poled in all the porous PVDF samples, indicating the piezoelectric origin of measured electrical charge from the applied force.

It is known that among the five crystalline polymorphs in PVDF, namely α, β, γ, δ and ε, the α and ε phases are non-polar with no electroactive response [43], while the ferroelectric properties are related to the polar phases of β, γ, and δ. β-PVDF is the most desired phase for ferroelectric applications since it exhibits the highest piezoelectric properties of the available PVDF polymorphs, with β-rich PVDF exhibiting a $d_{33}$ ~ -30 pC/N [44], compared to a lower $d_{33}$ ~ -7 pC/N [45] in γ-rich PVDF due to the smaller dipole moment. Therefore, the ferroelectric γ-phase in the freeze cast materials in this work (Fig. 4), is unlikely to be the main contributor to the high piezoelectric response of the freeze cast porous PVDF since the $d_{33}$ values are typically well in excess of 7 pC/N. This indicates that the porous semi-crystallized PVDF fabricated by ice-templating were piezoelectric, due to the formation of the dipole-like structure via applying a high electric field normal to the freezing direction and pore direction. This is in contrast to electret materials where charge flow is produced by either the change in a dielectric gap or the variation of the overlapping area [46]. Therefore, it can be concluded that the freeze-cast PVDF for all the weight ratios were ferroelectret in nature and, as shown in Fig. 5C. After applying a high poling electric field normal to the freezing direction, the air inside the aligned pore channel were subject to electrical breakdown, thereby depositing the opposite charges on the surface of the pore channel to form the dipole-like structure. Due to the high compliance of the porous polymer material, the polarisation and dipole moment is changed on application of a mechanical stress, thereby leading to an electro-mechanical response.

Since the piezoelectric properties of the ferroelectret originate from charged pores, it is of interest to evaluate their surface potential decay with time. The piezoelectric coefficient ($d_{33}$) of all samples decreased rapidly from 0.5 to 6 days followed by a slow reduction in $d_{33}$ values from 5 to 6 days measured after corona poling, as shown in Fig. 5D. Normally the surface potential, or $d_{33}$ values, fall rapidly in the first week for ferroelectrets based on polydimethylsiloxane (PDMS) [47], PDMS/Polyvinyl alcohol (PVA) [48], ethylene vinyl acetate copolymer (EVA) / biaxially oriented polypropylene (BOPP) [13], and fluorinated ethylene propylene (FEP)/polytetrafluoroethylene (PTFE) [49], then reach a more stable value, which can last for at least six weeks using porous polyethylene terephthalate (PET)/ethylene vinyl acetate copolymer (EVA) [50]. High charge capturing ability and low elastic modulus of the porous polymer have been considered as the main reasons for the stabilisation of the piezoelectric response in the ambient atmosphere with time [50]. In addition, the polymer nature and electrical properties, the ratio between the width and length of the elongated pore channel, and the thickness of the sample also play a crucial role in the change in piezoelectric performances of ferroelectrets with time.

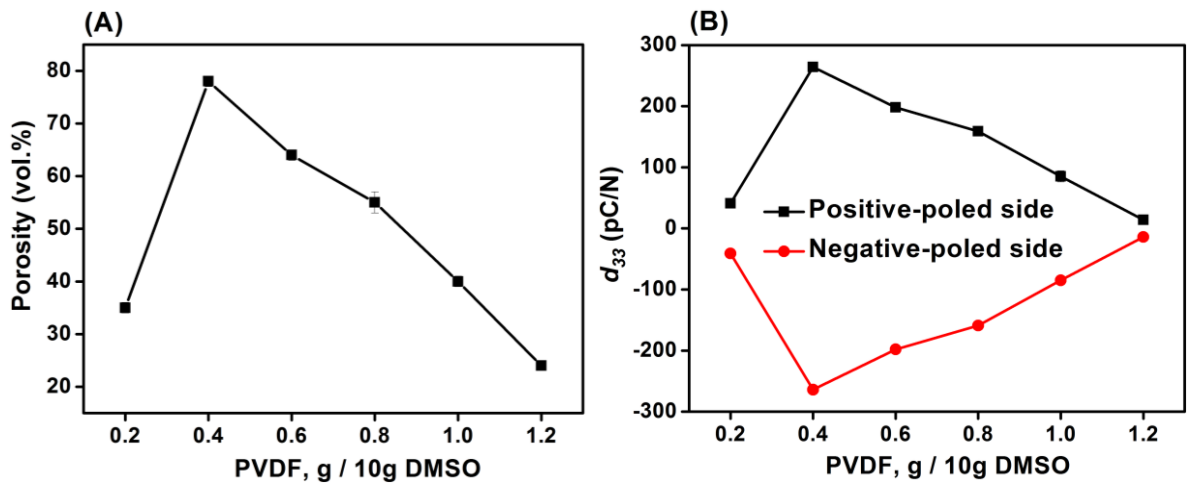

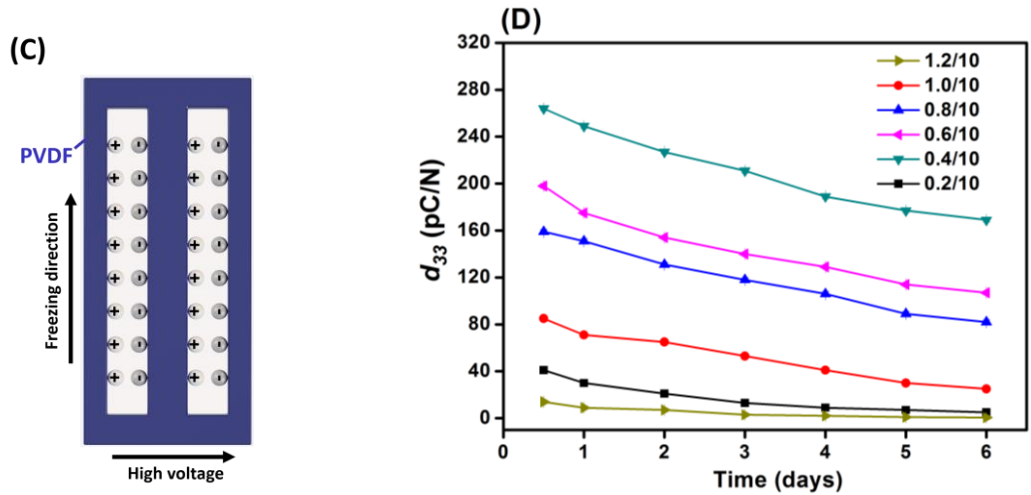

Figure 5: (A) porosity, (B) piezoelectric coefficient ($d_{33}$), (C) schematic of the plasma discharges on the surface and (D) aging performance of the porous PVDF with different weight ratios of PVDF/DMSO.

Figure 6 shows the dielectric properties of the porous PVDF measured from 1 Hz to 1MHz with different weight ratios ranging from 0.2/10 to 1.2/10, respectively. The relative permittivity of all the porous PVDF samples exhibited a frequency dependence at low frequency (~30 Hz) and a more frequency independent behaviour at higher frequencies (> 1 kHz, Fig. 6A), which is a common behaviour for disordered materials and which obeys the universal law of dielectric response [51]. The low frequency dispersion in permittivity is likely to be due to a small amount of conductivity in the sample [52] and this observation agrees with the higher dielectric loss (loss tangent) in the materials at low frequency (Fig. 6B). If the permittivity at the same frequency of 1 kHz is examined for all the samples (inset of Fig. 6A), the relative permittivity initially decreased from 3.0 at 0.2/10 until it reached a minimum of 1.5 at a weight ratio of 0.4/10, it then increased to 3.5, as the ratio increased to 1.2/10. Based on the microstructural observations, Fig. 2, and porosity results in Tab.1, the increase in porosity of PVDF from 35% to 78% as the weight ratios increased from 0.2/10 to 0.4/10 led to a reduction in permittivity of the porous PVDF and the subsequent increase relative permittivity was a result of the porosity decrease from 64% to 24% as the weight ratios increased from 0.6/10 to 1.2/10 (Fig. 6A). The dielectric loss (Fig. 6B) also revealed that at the same frequency the change of the dielectric loss varied in accordance with the change of the relative permittivity and porosity; see inset of Fig. 6A at 1 kHz. The combination of low permittivity and high piezoelectric activity can lead to high sensor sensitivity due to high $g_{33}$ (=$d_{33}/\varepsilon_{33}^\sigma$) coefficients, e.g. 19.9 Vm/N of 0.4/10 porous PVDF at 1 kHz compared with that of 0.067 Vm/N in the piezoelectric porous lead zirconate titanate (PZT) ceramic [53].

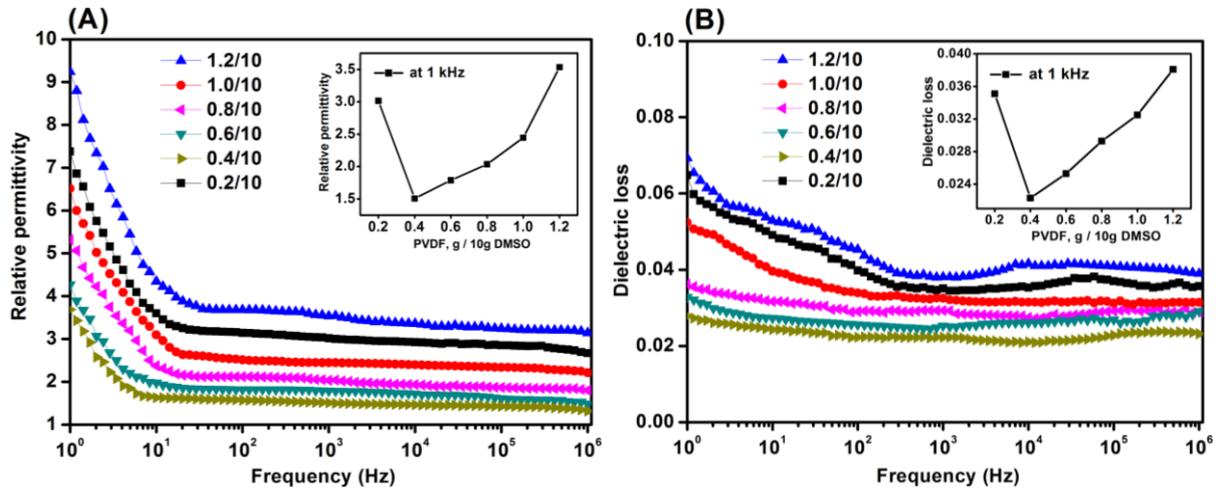

Figure 6: (A) Relative permittivity and (B) dielectric loss of the porous PVDF with different weight ratios of PVDF/DMSO.

**Conclusions**

In contrast to conventional processing methods to prepare ferroelectret materials with a cellular pore morphology, we have developed a new strategy to fabricate PVDF ferroelectret materials using an ice-templating method. The ferroelectret polymers are foams with charged pores and have the potential to be used in flexible wearable electronics, sensors and energy harvesters. The generated internal quasi-permanent macro-dipoles in the pore void can be induced by Paschen breakdown by application of a high electric field, where leads to piezoelectric properties. The ice-templated porous PVDF polymers formed by this new processing route exhibited elongated pore channels which were beneficial to achieve a high ferroelectret response. A maximum piezoelectric $d_{33}$ coefficient of ~264 pC/N was obtained for a porous ferroelectret PVDF with a pore size of 25 μm, which is significantly higher than a $d_{33}$ of 30 pC/N for dense β-phase ferroelectric PVDF films. In conclusion, this is the first demonstration of freeze casting in terms of its reliability and stability in the formation of ferroelectret polymers with highly aligned pore structures. Future avenues can include further optimisation of pore geometry or exploitation of the technique on polymer systems with longer charge and thermal stability, along with energy generation and sensor performance for practical flexible transducer applications.

**Acknowledgements**

Dr. Y. Zhang acknowledges support from the European Union's Horizon 2020 research and innovation programme under the Marie Skłodowska-Curie Grant, Agreement No. 703950 (H2020-MSCA-IF-2015-EF-703950-HEAPPs). Prof. C. R. Bowen, would like to acknowledge the funding from the European Research Council under the European Union's Seventh Framework Programme (FP/2007–2013)/ERC Grant Agreement No. 320963 on Novel Energy Materials, Engineering Science and Integrated Systems (NEMESIS). The authors are thankful for the helpful advice and inspiring discussions with Mr Sujoy Kumar Ghosh from Jadavpur University, India.